\documentclass[%
 aip,
 amsmath,amssymb,
 reprint,%
]{revtex4-1}

\usepackage{graphicx}
\usepackage{dcolumn}
\usepackage{bm}
\usepackage{scalerel}
\usepackage[utf8]{inputenc}
\usepackage[T1]{fontenc}
\usepackage{mathptmx}
\usepackage{xcolor}

\begin{document}

\preprint{AIP/123-QED}

\title[Hydrodynamic interactions between spheroids]{Hydrodynamic interactions and  the diffusivity of spheroidal particles\\}

\author{Navaneeth K. Marath}
\affiliation{%
Nordita, Royal Institute of Technology \& Stockholm University, Stockholm 106 91, Sweden
}%

\author{John S. Wettlaufer}
\affiliation{%
Yale University, New Haven, Connecticut 06520,United States
}%
\affiliation{%
Nordita, Royal Institute of Technology \& Stockholm University, Stockholm 106 91, Sweden
}%
\date{\today}

\begin{abstract}

It is intuitive that the diffusivity of an isolated particle differs from those in a monodisperse suspension, in which hydrodynamic interactions between the particles are operative.  Batchelor\cite{batchelor1976brownian,batchelor1983diffusion} calculated how hydrodynamic interactions influenced the diffusivity of a dilute suspension of spherical particles and Russel et al.,\cite{russel1991colloidal} and Brady\cite{brady1994long} treated non-dilute (higher particle volume fraction) suspensions.  
Although most particles lack perfect sphericity, little is known about the effects of hydrodynamic interactions on the diffusivity of spheroidal particles, which are the simplest shapes that can be used to model anisotropic particles. Here, we calculate the effects of hydrodynamic interactions on the translational and rotational diffusivities of spheroidal particles of arbitrary aspect ratio, in dilute monodisperse suspensions.  We find that the translational and rotational diffusivities of prolate spheroids is more sensitive to eccentricity than for oblate spheroids.  \textcolor{black}{The origin of the hydrodynamic anisotropy is that found in the stresslet field for the induced-dipole induced-dipole interaction.  However, in the dilute limit the anisotropy effects are at the level of a few percent.}  These effects have influence in a vast range of settings, from partially frozen colloidal suspensions to the dynamics of cytoplasm.  

\end{abstract}

\maketitle


\section{\label{sec:level1}Introduction\\}

Confinement of single particles and particle-particle interactions in suspensions, share a common set of hydrodynamic phenomena through which the single particle Stokes-Einstein diffusivity is modified.  The modification arises from the hydrodynamic interactions between suspended particles and/or particles in confined geometries.  Central are the effects of the particle-generated disturbance velocity field on the motion of other particles\cite{batchelor1976brownian}.  For spherical particles, the influence of hydrodynamic interactions is known for a range of the particle volume fractions \cite{russel1991colloidal}. However, little is known about the diffusivity of anisotropic particles, in no small part due to the difficulty in modeling the inter-particle interactions.  

This commonality of these hydrodynamic interactions naturally implies the existence of a swath of settings where they play out, ranging from confinement of single  Brownian\cite{FAUCHEUX:1994rt} and biopolymers\cite{Braun:2004yq}, to suspensions in experimental cells\cite{Brenner:1999xr} or proximity of particles due to phase changes of the solvent\cite{JSW2019} and subsequent restricted mobility\cite{Peppin:2009fe}.  In this latter setting, freezing of colloidal suspensions is an important phenomenon that  influences  many natural \cite{taber1929frost,hallet1990self} and technological \cite{mazur1970cryobiology} processes. The phase-change boundary in a freezing colloidal suspension can become morphologically unstable due to the constitutional supercooling of the suspension close to the boundary \cite{peppin2006solidification,peppin2006morphological,peppin2008experimental,MGW2018}. The degree of constitutional supercooling is influenced by the diffusivity of particles in the suspension, which affects the transport of the particles away from the boundary. The diffusivity also plays a role in determining the thickness of ice lenses\cite{schollick2016segregated}, and in the redistribution and aggregation of particles during the freezing of colloidal suspensions \cite{spannuth2011dynamics}.

\subsection{\label{sec:spheres } Spherical suspensions\\}

Prior to treating the diffusivity of spheroids, for context we briefly survey the effects of hydrodynamic interactions on the diffusivity of spherical particles.  
Dynamical light scattering experiments and the diffusivity of particles are intimately wed \cite{rallison1986effect,berne2000dynamic} and hence the key observations arise from this approach.  The mean squared displacement of a single Brownian spherical particle of radius $a$ and mass $m^s$ suspended in a Newtonian fluid of viscosity $\mu$ grows diffusively for times, $t\gg t^s_{\text{m}}=\frac{m^s}{6\pi\mu a}$,  where $t^s_{m}$ is the time scale for the relaxation of the momentum of the particle. The diffusivity of the particle is given by the Stokes-Einstein formula; $\mathbf{D}=\frac{k_b T}{6\pi \mu a}\mathbf{I}$, where $\mathbf{I}$ is the second-order isotropic identity tensor, $k_b$ is the  Boltzmann constant, and $T$  the temperature of the fluid. 
Batchelor\cite{batchelor1976brownian,batchelor1983diffusion} 
calculated the effects of hydrodynamic interactions on the diffusivity of a dilute suspension of spherical particles, which act as a leading order 
correction in the particle volume fraction ($\phi_p$) to the Stokes-Einstein diffusivity in the Stokes limit--when the effect of inertial forces in the fluid are neglected. For any suspension, one can define two kinds of diffusivity; the {\em gradient diffusivity}, which quantifies the flux of particles down their concentration gradient, and the {\em tracer diffusivity}, which relates the mean squared displacement of a tracer particle in the suspension to the elapsed time\cite{rallison1986effect}. Batchelor showed that due to hydrodynamic interactions the gradient (tracer) diffusivity for the suspension increases (decreases) relative to the single particle diffusivity.  Moreover, the tracer diffusivity takes on two values; one for short-times, $ t^s_{m} \ll t \ll t^s_{c}=a^2\left(\frac{k_b T}{6\pi \mu a}\right)^{-1}$, and another for long-times, $t\gg t^s_{c}$, which is the time needed for a particle in the suspension to diffuse its own distance.  Therefore, $t^s_{c}$ corresponds to the time scale at which there is a change in the configuration of the suspension. Batchelor's results were subsequently verified experimentally\cite{de1987sterically,russel1991colloidal}, using a Langevin approach \cite{pusey1983hydrodynamic} and using Stokesian dynamics simulations \cite{phillips1988hydrodynamic,brady1988stokesian}. In the non-dilute limit, that is with increasing particle volume fraction, the long-time tracer diffusivity decreases monotonically \cite{brady1994long}, whereas the gradient diffusivity does not\cite{russel1991colloidal,peppin2006morphological}.

\subsection{\label{sec:spheroids } Spheroidal suspensions\\}

Here, we investigate the effects of hydrodynamic interactions on the diffusivity of anisotropic particles by modeling them as spheroids, which are axisymmetric ellipsoids, whose shapes are determined by their aspect ratio. The aspect ratio ($r$) is the ratio of a spheroid's length along its axisymmetric axis to its maximum diameter perpendicular to the axis, and is thus a measure of its anisotropy. Spheroids with $r<1$ and $r>1$ are called oblate and prolate spheroids respectively; a sphere has $r=1$, a flat disk has $r \rightarrow 0$ and a slender fiber $r \rightarrow \infty$.

\begin{figure}
\includegraphics[scale=.7]{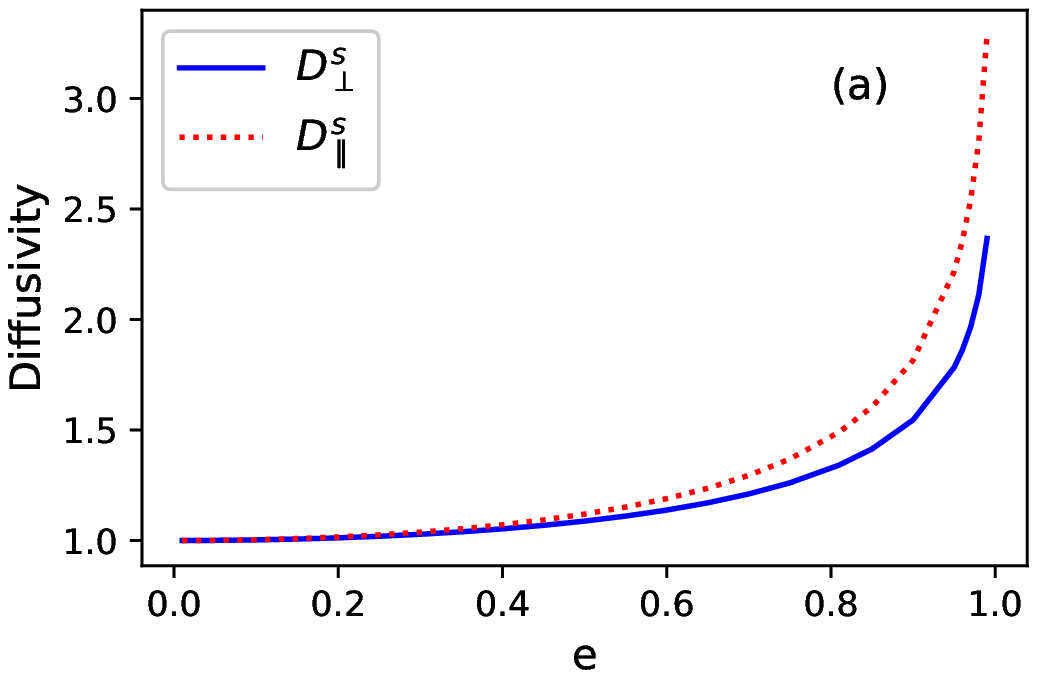}\\
\includegraphics[scale=.7]{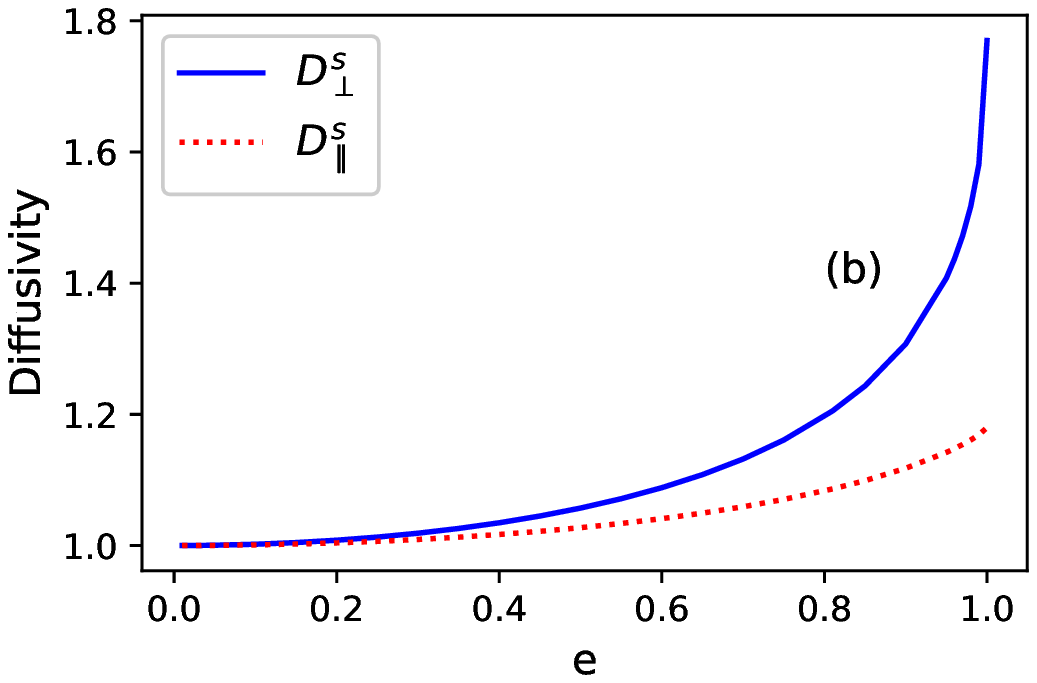}
\caption{\label{figuretransstokes} The dimensionless (scaled by $\frac{k_b T}{6\pi \mu L}$) translational diffusivities parallel (dashed) and perpendicular (solid) to the axisymmetric axis of a spheroid versus eccentricity, for prolate (a) and oblate (b) spheroids. }
\end{figure}

\begin{figure}
\includegraphics[scale=.7]{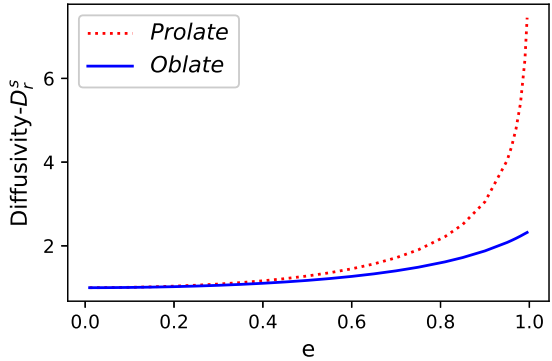}\\
\caption{\label{figurerotstokes} The dimensionless rotational diffusivity (scaled with $\frac{k_b T}{8\pi \mu L^3}$, transverse to the axis  of a spheroid) versus eccentricity.}
\end{figure}

A single spheroid executes diffusive motion for times, $t\gg t_{m}= \max\left[\frac{m}{6\pi\mu L \,  \min(X_A,Y_A)}, \frac{I_t}{8\pi\mu L^3 Y_c}\right]$, where $L$ is the length of the semi-major axis of its generating ellipse, $m$ is its mass, $I_t$ is its moment of inertia transverse to the axis, $X_A$, $Y_A$ and $Y_C$ are functions of the eccentricity\cite{kim2013microhydrodynamics}, $6\pi\mu LX_A$ ($6\pi\mu LY_A$) is the translational drag along  (perpendicular) to the axis, and $8\pi\mu L^3 Y_c$ is the rotational drag torque transverse to the axis. The eccentricity of an oblate (prolate) spheroid is defined as $e=\sqrt{1-r^2}$ ($e=\sqrt{1-1/r^2}$), ranging from $0$ to $1$ as the shape ranges from a sphere to a slender fiber or a flat disk.  
The largest of the translational and the angular momentum relaxation time scales is given by $t_{m}$ above. The translational ($\mathbf{D_{tr}}$)  and rotational (${D_r}$) diffusivities of a neutrally buoyant spheroid suspended in a Newtonian fluid of dynamic viscosity $\mu$ are given by
\begin{eqnarray}
\mathbf{D_{tr}}=\displaystyle \frac{k_b T}{6\pi\mu L}\left[\frac{1}{X_A}\, \mathbf{p}_1 \mathbf{p}_1 +  \frac{1}{Y_A} (\mathbf{I} -\mathbf{p}_1 \mathbf{p}_1) \right]\label{eq:transspheroiddiffusivity}
\end{eqnarray}
and 
\begin{eqnarray}
{D_r}=\displaystyle \frac{k_b T}{8\pi\mu L^3} \frac{1}{Y_C}\,\label{eq:rotspheroiddiffusivity}
\end{eqnarray}
respectively, and are shown as a function of the spheroid eccentricity in figures \ref{figuretransstokes} and \ref{figurerotstokes}, where $D_{\parallel}^{s}=X_A^{-1}$, $D_{\perp}^{s}=Y_A^{-1}$ and $D_{r}^{s}=Y_C^{-1}$.  The orientation of the spheroid is given by $\mathbf{p}_1$, and $\mathbf{D_{tr}}$ is anisotropic for $X_A\neq Y_A$.  The translational diffusivities parallel and perpendicular to $\mathbf{p}_1$, reduce to the Stokes-Einstein diffusivity, when $e=0$. In the limit that $e\rightarrow 1$, the diffusivities diverge as $\log{(1-e)}$ for prolate spheroids, and asymptote to finite values ($D_{\perp}^{s}=1.77$ and $D_{\parallel}^{s}=1.18$) for oblate spheroids. Furthermore, when $e \neq 0$ the diffusivity parallel to the axis is larger (smaller) than that perpendicular to the axis for prolate (oblate) spheroids. Figure \ref{figurerotstokes} shows that ${D_r}(e=0)$ = $\frac{k_b T}{8\pi \mu L^3}$,  the spherical value with dimension $s^{-1}$, and ${D_r}(e\rightarrow 1)$ diverges as $\log{(1-e)}$ for the prolate spheroid, and asymptotes to a finite value ($D_r^s=2.35$) for the oblate spheroid. 
On a time scale $t \gg D_r^{-1}$, the translational diffusivity of a spheroid given in Eq.~(\ref{eq:transspheroiddiffusivity}) will become isotropic owing to this rotational diffusion, and thus unlike spherical suspensions, even for a single spheroid, there are two translational diffusivities, one at short-times ($t \ll D_r^{-1}$) and the other at long-times ($t \gg D_r^{-1}$). 

Hydrodynamic interactions insure that both the gradient and tracer diffusivities of a spheroidal suspension differ from those given in Eqs.~(\ref{eq:transspheroiddiffusivity}) and (\ref{eq:rotspheroiddiffusivity}).
Importantly, the tracer diffusivity should take different forms: (a) For short-times, $t_{m}\ll t \ll t_{c} =\min\left[D_r^{-1}, L^2\left(\frac{k_b T}{6\pi \mu L \min({X_A,Y_A})}\right)^{-1}\right]$, and (b) for long-times, $t\gg \max\left[D_r^{-1}, L^2\left(\frac{k_b T}{6\pi \mu L \min({X_A,Y_A})}\right)^{-1}\right]$. Here, $t_c$ is the time scale at which the configuration of the suspension changes. Before describing our calculation, we survey previous investigations of the diffusivity of spheroidal particles in monodisperse suspensions. 

Treloar and Masters \cite{treloar1991short} calculated the effects of hydrodynamic interactions on the short-time translational diffusivity of {\em nearly spherical} spheroids ($e\ll1$), and hypothesized that the diffusivity for small $e$, might be a good approximation for all values of $e$. Claeys and Brady \cite{claeys1993suspensions2} used Stokesian dynamics to calculate the short-time translational and rotational diffusivities of a prolate spheroid with aspect ratio $6$ in unbounded monodisperse suspensions, 
and found that the diffusivity decreased monotonically with increasing particle volume fraction.  
Zheng and Han\cite{zheng2010self} measured the translational diffusivity of a tracer spheroid in a monolayer of spheroids ($r=9.07$) near a flat wall (a two-dimensional suspension) as a function of spheroid area fraction.  They found that as the particle area fraction increases, the anisotropy of the diffusivity decreases, reaching a minimum.   Upon further increase in area fraction the anisotropy of the diffusivity increased.  However, their system is two-dimensional and their results cannot be translated to unbounded three-dimensional suspensions.

In this paper, we calculate the short-time translational and rotational diffusivities of a tracer spheroid of an arbitrary aspect ratio in a dilute monodisperse suspension. The correction to the single particle diffusivities appears at O($\phi$), where $\phi$ is the hydrodynamic volume fraction defined as $nL^3$, with $n$ the number density of the spheroids in the suspension. We use a far-field approximation to model the hydrodynamic interactions, and a Langevin approach to calculate the diffusivities.  In section~\ref{sec:analysis}, we describe the calculation of the Langevin equation for $N$ spheroids to obtain the mean squared angular and translational displacements of one spheroid, and the diffusivities follow from averaging  over the configuration space of the other spheroids. The results are discussed in section~\ref{sec:Results} before concluding in section~\ref{sec:Conclusions}.  To facilitate continuity in the presentation four Appendixes contain a number of technical details.  

\section{\label{sec:analysis}Analysis\\}
Here we calculate the short-time diffusivities of a tracer spheroid in a monodisperse suspension of $N$ spheroids using a Langevin equation approach as follows. The diffusivities are related to the mean squared rotational and angular displacements of the tracer spheroid at time $t$, with  $t_{m}\ll t \ll t_{c}$. 
The mean squared displacements are obtained by solving the Langevin equation for $N$ spheroids (see Ref.~(\onlinecite{ermak1978brownian}) for the case of $N$ spheres). For each spheroid in the suspension, we define a Cartesian coordinate system {$x_{i}$, $ y_{i}$, $ z_{i}$, with $i=1,\dots N$\} such that the orientation of the spheroid, ${\mathbf{p}_i}$, is aligned with the $z_i$ axis. The coordinate system does not rotate with the spheroid. Noting that the moments of inertia of any spheroid along the transverse axes ($x_{i}$ and $y_i$) are the same, one can write down the Langevin equation in index notation as     
\begin{eqnarray}
\,{M}_{ij} {\dot{X}}_j-  (I_s-I_t) X_i^{I}= -k_{b} T\, D_{ij}^{-1} X_j  +   {\alpha}_{ij} g_j, \label{eq:langeq}
\end{eqnarray}
where the terms on the left-hand side arise due to the inertia of the spheroids, and the first and the second terms on the right-hand side are the deterministic and the stochastic forces exerted by the fluid on the spheroids respectively. \textcolor{black}{In Eq.~(\ref{eq:langeq}), we have neglected all the direct forces between the spheroids}.

In Eq.~(\ref{eq:langeq}), $\mathbf{M}$ is a generalized mass/moment of inertia diagonal matrix of dimension $6N\times6N$ , with $M_{ii}= m$ for $i=1 $ to $3N$, $M_{ii}= I_t$ for $i=3N+j,3N+2j$ and $M_{ii}= I_s$ for $i=3N+3j$, with $j=1 $ to $N$. The constants $I_s$ and $I_t$ are the components of the moment of inertia of any of the $N$ spheroids along and transverse to its orientation respectively, and $m$ is the spheroid mass. Additionally, $\mathbf{X}$ and $\mathbf{g}$ are $6N$-dimensional vectors.  The first vector contains the components of the translational ($v$) and the angular ($\omega$) velocities of the $N$ spheroids along the $x_i,y_i$ and $z_i$ axes, and the second vector contains the components of the translational ($F$) and the rotational ($\tau$) Brownian noises acting on the $N$ spheroids along these axes, and are given by
\begin{eqnarray}
\mathbf{X}=\begin{bmatrix}
    {v}_{x_1},{v}_{y_1},  \hdots ,{v}_{z_N}, {\omega}_{x_1}, {\omega}_{y_1}, {\omega}_{z_1},\hdots,  {\omega}_{y_N},{\omega}_{z_N}\end{bmatrix}^{T} \nonumber,
\end{eqnarray}
and
\begin{eqnarray}
\mathbf{g}=\begin{bmatrix}
    {F}_{x_1},{F}_{y_1},  \hdots ,{F}_{z_N}, {\tau}_{x_1}, {\tau}_{y_1}, {\tau}_{z_1},\hdots,  {\tau}_{y_N},{\tau}_{z_N}\end{bmatrix}^{T} \nonumber.
\end{eqnarray}
In Eq.~(\ref{eq:langeq}), the term proportional to $\mathbf{X}^I$ is a $6N$-dimensional vector that describes the rate of change of the moment of inertia of the spheroid along the $x_i,y_i$ and $z_i$ axes, on account of the rotation of the spheroid\cite{goldstein2011classical}.  We have ${X}^I_i=0$, for $i=1 $ to $3N$,  ${X}^I_i=-{\omega}_{y_j}{\omega}_{z_j}$, for $i=3N+j$, ${X}^I_i={\omega}_{x_j}{\omega}_{z_j}$, for  $i=3N+2j$ and ${X}^I_i= 0$, for $i=3N+3j$, with $j=1 $ to $N$. Note that the rate of change vanishes along the symmetry axis of the spheroid, and for a sphere, $\mathbf{X}^I$ is a null vector. We assume that the noise terms have zero mean,  are Gaussian and $\delta$-autocorrelated in time as
\begin{eqnarray}
\langle g_i(t) g_j(t') \rangle =2 \delta_{ij} \, \delta (t-t')\label{eq:forcecorr}.
\end{eqnarray}
The angular brackets above denote an ensemble average over the rapidly fluctuating random forces and torques for a given configuration of $N$ spheroids. The quantity {$(k_b T)^{-1} \mathbf{D}$ is the mobility matrix for $N$ spheroids, relating their translational and the angular velocities to the hydrodynamic torques ($\tau^h$) and forces ($F^h$) acting on them as
\begin{eqnarray}
\mathbf{X}=&(k_b T)^{-1} \mathbf{D}\cdot\mathbf{\mathcal{F}}^h \label{eq:mob1} ,
\end{eqnarray}
where
\begin{eqnarray}
\mathbf{\mathcal{F}}^h =&
\begin{bmatrix}
    {F}_{x_1}^h,{F}_{y_1}^h,  \hdots ,{F}_{z_N}^h, {\tau}_{x_1}^h, {\tau}_{y_1}^h, {\tau}_{z_1}^h,\hdots,  {\tau}_{y_N}^h,{\tau}_{z_N}^h\end{bmatrix}^{T},
\end{eqnarray}
and
\begin{eqnarray}
\mathbf{D} =&
\begin{bmatrix}
    {D}_{\scaleto{11}{3pt}}^{tt} &{D}_{\scaleto{12}{3pt}}^{tt}&{D}_{\scaleto{13}{3pt}}^{tt} &. &D_{\scaleto{13N}{3pt}}^{tt}&{D}_{\scaleto{11}{3pt}}^{tr}&{D}_{\scaleto{12}{3pt}}^{tr}&.& {D}_{\scaleto{13N}{3pt}}^{tr}\\
    {D}_{\scaleto{21}{3pt}}^{tt} &{D}_{\scaleto{22}{3pt}}^{tt}&{D}_{\scaleto{23}{3pt}}^{tt} &. &D_{\scaleto{23N}{3pt}}^{tt}&{D}_{\scaleto{21}{3pt}}^{tr}&{D}_{\scaleto{22}{3pt}}^{tr}&.& {D}_{\scaleto{23N}{3pt}}^{tr}\\
  \vdots & \vdots &\vdots &\vdots &\vdots &\vdots &\vdots &\vdots &\vdots\\
  {D}_{\scaleto{3N1}{3pt}}^{tt}&. &. &. &{D}_{\scaleto{3N3N}{3pt}}^{tt}&{D}_{\scaleto{3N1}{3pt}}^{tr}&{D}_{\scaleto{3N2}{3pt}}^{tr}&.& {D}_{\scaleto{3N3N}{3pt}}^{tr}\\
  {D}_{\scaleto{11}{3pt}}^{rt}&. &. &. &{D}_{\scaleto{13N}{3pt}}^{rt}&{D}_{\scaleto{11}{3pt}}^{rr}&{D}_{\scaleto{12}{3pt}}^{rr}&.& {D}_{\scaleto{13N}{3pt}}^{rr}\\
  \vdots&\vdots  & \vdots &\vdots &\vdots &\vdots &\vdots&\vdots&\vdots \\
  {D}_{\scaleto{3N1}{3pt}}^{rt}&.&.  &. &{D}_{\scaleto{3N3N}{3pt}}^{rt}&{D}_{\scaleto{3N1}{3pt}}^{rr}&{D}_{\scaleto{3N2}{3pt}}^{rr}&.& {D}_{\scaleto{3N3N}{3pt}}^{rr}\\\end{bmatrix}.
\end{eqnarray}
The two superscripts of a matrix element indicate whether it multiplies a force (`$t$') or a torque (`$r$') component, to give a translational (`$t$') or  an angular (`$r$') velocity component in Eq.~(\ref{eq:mob1}). The mobility matrix is only a function of the configuration--positions and orientations--of the $N$ spheroids. In Eq.~(\ref{eq:langeq}), 
$\boldsymbol{\alpha}$ is the strength of the stochastic noise, which is related to the mobility matrix through the fluctuation-dissipation theorem given by
\begin{eqnarray}
(k_b T)^2 D_{ij}^{-1} = \alpha_{ik}\alpha_{jk}, \label{eq:flucdissi}
\end{eqnarray}
which is derived in Appendix \ref{app:fdt}. Therefore, the strength of the fluctuations depends on the configuration of the spheroids through Eq.~(\ref{eq:flucdissi}), so that 
 the positions and the orientations of the spheroids are influenced by  multiplicative noise.
 
In Appendix \ref{app:subsec2}, Eq.~(\ref{eq:langeq}) is solved using Eqs.~(\ref{eq:forcecorr}) and (\ref{eq:flucdissi}) to obtain the mean squared translational and rotational displacements of a tracer spheroid (say 1), which are given by   
\begin{eqnarray}
\langle\Delta r_p^1  \Delta r_q^1 \rangle= 2 \langle D^{tt}_{pq} \rangle t,\label{eq:msdtrans}
\end{eqnarray}
and 
\begin{eqnarray}
\langle\Delta \theta_p^1  \Delta \theta_q^1 \rangle= 2 \langle D^{rr}_{pq} \rangle t\label{eq:msdrot}.
\end{eqnarray}
In  Eqs.~(\ref{eq:msdtrans}) and (\ref{eq:msdrot}), $\Delta r_p^1$ and $\Delta \theta_p^1$ are the translational and the rotational displacements of the tracer spheroid `1' from its initial position/orientation at $t=0$, with $p,q={1,2 \mbox{ and } 3}$.
 The angular brackets denote the ensemble average over all possible configurations of the other spheroids in the suspension, with the tracer spheroid fixed in space.

The mean squared displacements in Eqs.~(\ref{eq:msdtrans}) and (\ref{eq:msdrot}) are valid even in the non-dilute limit. However, in the dilute limit, they can be simplified further by neglecting the simultaneous interactions between three or more spheroids, since these interactions lead to corrections that are less than or equal to O($\phi^2$), as seen in the case for spheres \cite{dhont1996introduction}. Because the mobility matrix is pairwise additive \cite{durlofsky1987dynamic}, the averaging procedure reduces to that over the configuration space of two spheroids with the tracer spheroid fixed in space. In Appendix \ref{app:mobility} we use the method of reflections\cite{kim1985sedimentation, kim2013microhydrodynamics} to derive the mobility matrix for two prolate spheroids under the influence of forces and torques, 
 and we truncate the mobility matrix at the level of force dipoles, or stresslets, and neglect higher multipole moments. 
 mportantly, these higher multipole moments become as important as the lower moments when the interparticle distances is small compared to the particle length, in which case lubrication forces become operative and can be crucial in non-dilute suspensions.
 
The method of reflections must include at least two reflections to insure that the velocity (angular velocity) of the first spheroid has a non-trivial relation to 
the force (the torque) acting on it.   The trivial relation is given by Eq.~(\ref{eq:finmob1}) [Eq.~(\ref{eq:finmob2})] for the translational [angular] velocity, and the non-trivial relation given by Eq.~(\ref{eq:finmob3}) [Eq.~(\ref{eq:finmob4})]. The averaging of the trivial relations over the configuration space recovers the single particle diffusivities given in Eqs.~(\ref{eq:transspheroiddiffusivity})  and (\ref{eq:rotspheroiddiffusivity}). For averaging, we use only the leading order term in the non-trivial relation, which decays as $1/R_p^4$ ($1/R_p^6$) for translational (angular) velocity, where $R_p$ is the interparticle distance (non-dimensionalized by `$L$') between the centers of the spheroids. 

The mobility matrix for two oblate spheroids is obtained from that of two prolate spheroids by using the transformation described in Appendix \ref{app:mobility2}. The mobility matrices are four-dimensional integrals, and the averaging in Eqs.~(\ref{eq:msdtrans}) and 
(\ref{eq:msdrot}) requires an additional integration over a five-dimensional space; two for the orientation of the second spheroid and three for the interparticle vector connecting the centers of the spheroids. The nine-dimensional integrals required to average the diffusivities are given by
\begin{eqnarray}
\langle D^{tt}_{pq} \rangle  =\frac{1}{4\pi V}\int_{0}^{\pi}\int_0^{2 \pi}\int_0^{\pi}\int_0^{2 \pi} \int_{R_\text{restrict}}^{\infty}D^{tt}_{pq} R_p^2 \sin\theta_R\nonumber \\\sin\theta_2    dR_p d\phi_2 d\theta_R d\phi_R d\theta_2  \label{eq:finalinteg1},
\end{eqnarray}
and
\begin{eqnarray}
\langle D^{rr}_{pq} \rangle  =\frac{1}{4\pi V}\int_{0}^{\pi}\int_0^{2 \pi}\int_0^{\pi}\int_0^{2 \pi} \int_{R_\text{restrict}}^{\infty}D^{rr}_{pq} R_p^2 \sin\theta_R \nonumber \\   \sin\theta_2dR_p d\phi_2 d\theta_R d\phi_R d\theta_2 \label{eq:finalinteg2}.
\end{eqnarray}
In these integrals, all the configurations in which the second spheroid intersects the tracer spheroid are avoided, and $V$ corresponds to that volume in which one can place the second spheroid without intersecting the tracer spheroid. We non-dimensionalize all distances with `$L$', and the inter-particle separation vector $\mathbf{R_p}$ is defined as ($R_p$,$\theta_R$,$\phi_R$) in a spherical coordinate system whose zenith direction is aligned with the orientation of the tracer spheroid. Because we avoid intersections described above, the inner limit of $R_p$, defined as $R_\text{restrict}$, will depend on $\theta_R$ and $\phi_R$. The orientation of the second spheroid is defined in the spherical coordinate system as ($1$,$\theta_2$,$\phi_2$). We make a further simplification by setting $R_\text{restrict}=2$, which insures that the spheroids don't intersect. The nine-dimensional integrals are evaluated  numerically using a Monte Carlo method with a sampling size of $10^{9}$ configurations.  Additionally, we evaluated the integrals using the NIntegrate function in Mathematica as a cross check.

\section{\label{sec:Results}Results\\}

The short-time translational ($\mathbf{D^f_{tr}}$) and rotational (${D^f_{r}}$) diffusivities of the tracer spheroid that are obtained by numerically integrating the right-hand sides of Eqs.~(\ref{eq:finalinteg1}) and (\ref{eq:finalinteg2}) are written as
\begin{eqnarray}
 \mathbf{D^f_{tr}}=& \mathbf{D_{tr}} + \phi \, \displaystyle  \frac{k_b T}{6 \pi \mu L} \left[ D_{\parallel}  \mathbf{p_1p_1}+ D_{\perp}(\mathbf{I}-\mathbf{p_1p_1}) \right] \label{eq:finalanswer},
 \end{eqnarray}
 and
 \begin{eqnarray}
  {D^f_{r}}=& {D_{r}} + \phi \, \displaystyle   \frac{k_b T}{8 \pi \mu L^3} D_r^\text{hyd}, \label{eq:finalanswer2}
\end{eqnarray}
in which $\mathbf{D_{tr}}$ and ${D_{r}}$ are the single particle diffusivities given in Eqs.~(\ref{eq:transspheroiddiffusivity}) and (\ref{eq:rotspheroiddiffusivity}). In Eq.~(\ref{eq:finalanswer}), $D_{\parallel}$ and  $D_{\perp}$ are the O($\phi$) correction to the translational diffusivities parallel and perpendicular to the spheroid respectively, and $D_r^\text{hyd}$ in Eq.~(\ref{eq:finalanswer2}) is the O($\phi$) correction to the rotational diffusivity.

The corrections $D_{\parallel}$ and $D_{\perp}$ are plotted as a function of the eccentricity of prolate and oblate spheroids in figure \ref{figuretrans}. Because these  corrections are negative for all eccentricities, the hydrodynamic interactions reduce the translational diffusivity of the tracer spheroid. 
Physically, this reduction is due to the hydrodynamic hindrance created by the velocity fields of the other particles in the suspension, which on average suppress the displacement of the tracer spheroid. Spherical symmetry is exhibited in figures \ref{figuretrans} (a) and (b), where for $e=0$, $D_{\parallel}$ and  $D_{\perp}$ take the same value of $-1.87$ \footnote{Batchelor\cite{batchelor1976brownian} obtained a slightly different value of $-1.83$ for the O($\phi$) correction, by using a mobility matrix which was exact for any arbitrary separation between two spheres, whereas we use a mobility matrix which is approximate, since only two reflections are considered. However, even with the approximate mobility matrix, we obtain a good estimate for the correction at $e=0$.}. Both the corrections decrease \textcolor{black}{monotonically} with increasing eccentricity \textcolor{black}{because (a) the interactions weaken for a given $\phi$($nL^3=$  constant)}, and (b) in the limit of $e\rightarrow1$, vanish as $\propto 1/\log(1-e)$  for prolate spheroids, and for oblate spheroids, $D_{\parallel}$ and $D_{\perp}$ asymptote to $-0.34$ and  $-0.35$ respectively. Physically, in this limit the hydrodynamic interactions vanish for slender fibers whereas they survive for flat disks due to the finite area of the latter.
 For nearly spherical prolate spheroids ($e\ll1$), $D_{\parallel}$ and $D_{\perp}$ asymptote to $-1.87+2.00\,e^2$ and $-1.87+2.06\, e^2$ respectively, and for nearly spherical oblate spheroids, $D_{\parallel}$ and $D_{\perp}$ asymptote to $-1.87+0.81 e^2$  and $-1.87+0.75\,e^2$  respectively. Thus, we find that, as hypothesized by Treloar and Masters \cite{treloar1991short}, the nearly  spherical asymptotes are not a uniformly valid approximation for all eccentricities. 

 \textcolor{black}{ In figures \ref{figurepercentp} (a) and \ref{figurepercento} (a), the hydrodynamic corrections ($\phi D_{\parallel}$ and $\phi D_{\perp}$) for prolate and oblate spheroids are expressed as a percentage of their associated single particle diffusivities ($D_{\parallel}^s$ and $D_{\perp}^s$), and plotted as a function of their eccentricities for $\phi=0.1$. Clearly, the largest hydrodynamic reduction in diffusivity occurs for spheres, and for both prolate and oblate spheroids the reductions decrease monotonically as $e\rightarrow 1$.  We see that for eccentricities greater than approximately $0.4$, there is a small difference between the reductions in the parallel and perpendicular diffusivities; for prolate (oblate) spheroids, the reduction is slightly larger for the perpendicular (parallel) diffusivity than for the parallel (perpendicular) diffusivity. The physical origin of the difference can be described using the case of the prolate spheroid.  The hydrodynamic effect of a moving spheroid in the suspension is to induce stresslets (dipoles) on the other spheroids, which in turn induce the same back upon the spheroid. On average, because the gradient in the reflected velocity field is likely to be larger (smaller) parallel to (perpendicular to) the spheroidal axis, so too is the strength of the induced dipole\footnote{The net change in the reflected velocity field along a spheroidal axis is the integral over the spheroid of the gradient of the velocity field along that direction,  and it is averaged over all possible configurations of the suspension with the test spheroid fixed.  This underlies the strength of the induced dipole.}.  Now, the radial dipole flow field is towards (away from) the spheroid perpendicular to (parallel to) the axis \cite{kim2013microhydrodynamics}, thereby suppressing (enhancing) perpendicular (parallel) motion, as seen in figure \ref{figurepercentp} (a). This 
logic is immediately extended to the oblate spheroid, in which case the inward radial flow due to the induced dipole will be parallel to the spheroidal axis.  Finally, as the eccentricity increases the {\em anisotropy} of the dipole strength increases, whereas the {\em strength} decreases, due to the reduction in the size of the spheroid, and the latter effect dominates the behavior as $e\rightarrow1$.}

\textcolor{black}{To characterize the effects of the anistropy on $\mathbf{D^f_{tr}}$, we define a ratio of the diffusivities, $D_a$, as
 \begin{eqnarray}
 D_a=\frac{D_{\parallel}^s+ \phi D_{\parallel}}{D_{\perp}^s+ \phi D_{\perp}}, \label{eq:ratio}
 \end{eqnarray}
 which is plotted for prolate (oblate) spheroids versus eccentricity for various volume fractions in figures \ref{figurepercentp} (b) and \ref{figurepercento} (b). Clearly, there is a weak $\phi$-dependence, exhibited differently in figures \ref{figurepercentp} (c) and \ref{figurepercento} (c) by scaling  $D_a$ with its value at $\phi=0$.  Here we see that the maximum change in the ratio for prolate (oblate) spheroids is about $0.8\%$ ($2.5\%$) for the the largest volume fraction considered. Therefore, at O($\phi$), the hydrodynamic interactions do not significantly affect the anisotropy of the diffusivity tensor.
 }

In figure \ref{figurerot}, the orientational diffusivity correction $D_r^\text{hyd}$ of Eq.~(\ref{eq:finalanswer2}) is plotted versus the eccentricity of prolate and oblate spheroids. The correction is negative as in the translational case, and starts at the sphere value of $-0.31$ at $e=0$. The orientational diffusion does not affect the dynamics of a spherical particle due to its isotropy, and $D_r^\text{hyd}$ has a finite value just as in the single particle case (see figure \ref{figurerotstokes}).  For prolate spheroids, in the limit of $e\rightarrow1$, the correction decays as $1/\log(1-e)$ because of the weakened hydrodynamic interactions, and for oblate spheroids in the same limit the correction asymptotes to  $-0.04$. \textcolor{black}{In figure \ref{figurerot2}, the correction ($\phi D_r^\text{hyd}$) for a spheroid is expressed as a percentage of the associated single particle diffusivity ($D_r^s$), and plotted versus eccentricity. The percentage reduction in the orientational diffusivities of prolate and oblate spheroids are far smaller than that found for their translational counterparts in figures \ref{figurepercentp} (a) and \ref{figurepercento} (a)}.  

Claeys and Brady\cite{claeys1993suspensions2} used Stokesian dynamics to calculate the short-time diffusivities of a tracer spheroid of aspect ratio $6$ ($e=0.98$) in a monodisperse suspension. They modeled the hydrodynamic interactions by including lubrication forces, in addition to a far-field approximation of the mobility matrix. When averaging the diffusivity, they generated different configurations of $N$ spheroids (a maximum $N = 64$) in a unit cell, with periodic boundary conditions, using a Monte Carlo procedure. They reported a linear combination of the translational diffusivities parallel and perpendicular to the symmetry axis of the tracer spheroid, which decreased with increasing volume fraction. 
\textcolor{black}{ In particular, for small volume fractions, their result is in excellent agreement with ours for the linear combination; $2.51-0.07\phi$ (when scaled with $\frac{k_b T}{6\pi\mu L}$) from  figures \ref{figuretransstokes} (a) and \ref{figuretrans} (a). }

\begin{figure}
\includegraphics[scale=.7]{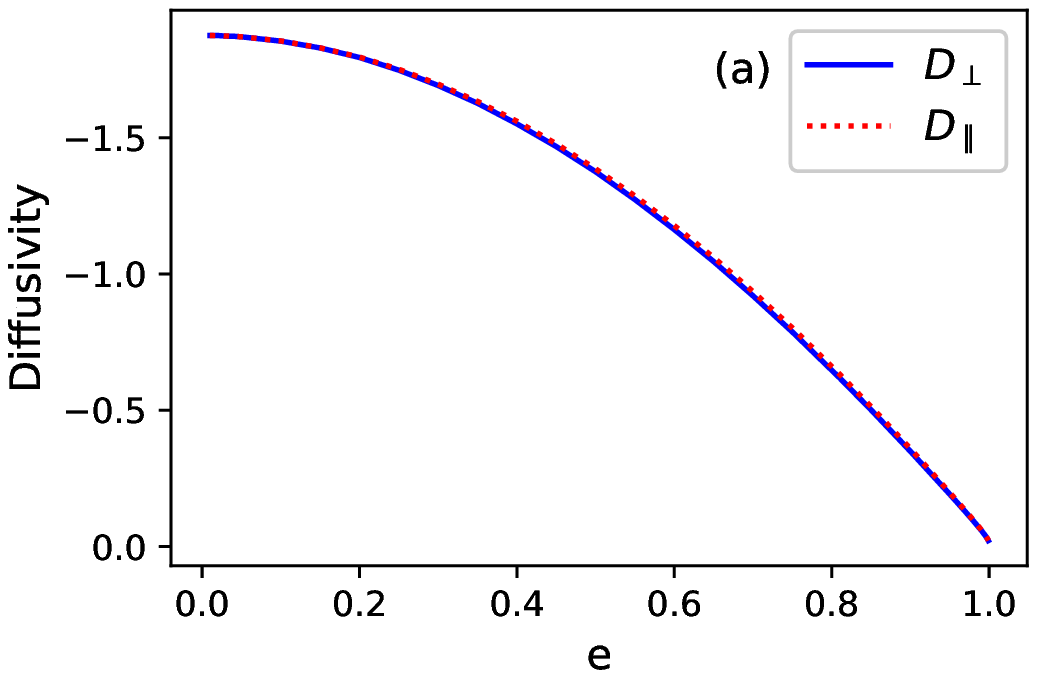}\\
\includegraphics[scale=.7]{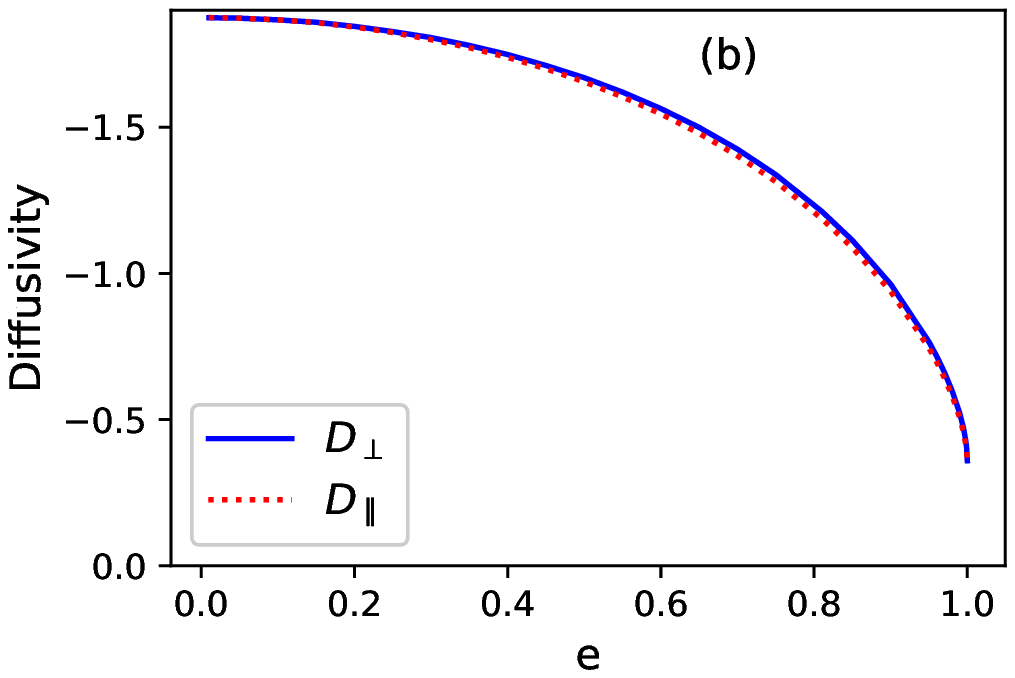}
\caption{\label{figuretrans} The O($\phi$) hydrodynamic corrections to the translational diffusivity given in Eq.~(\ref{eq:finalanswer}) versus the eccentricity of (a) prolate and (b) oblate spheroids. }
\end{figure}

\begin{figure}
\includegraphics[scale=.7]{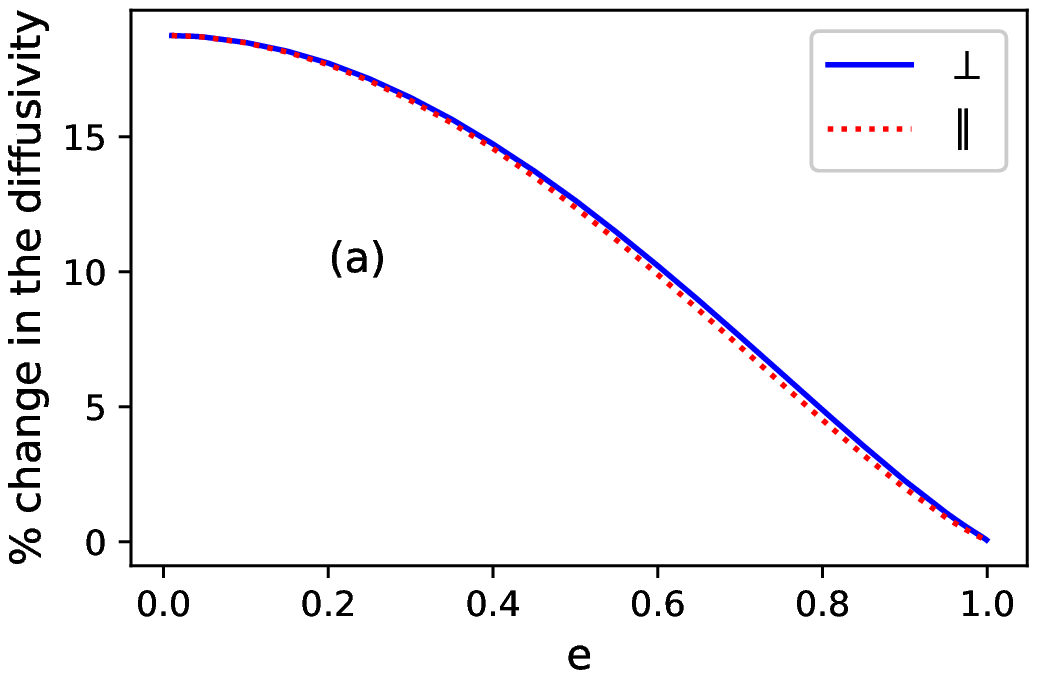}\\
\includegraphics[scale=.7]{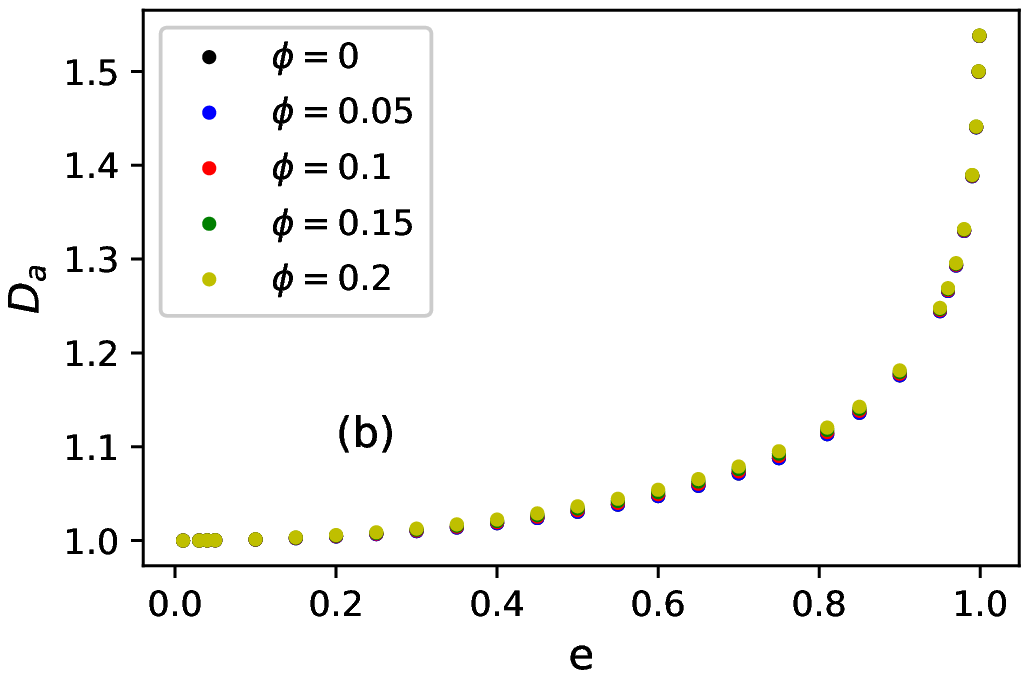}\\
\includegraphics[scale=.7]{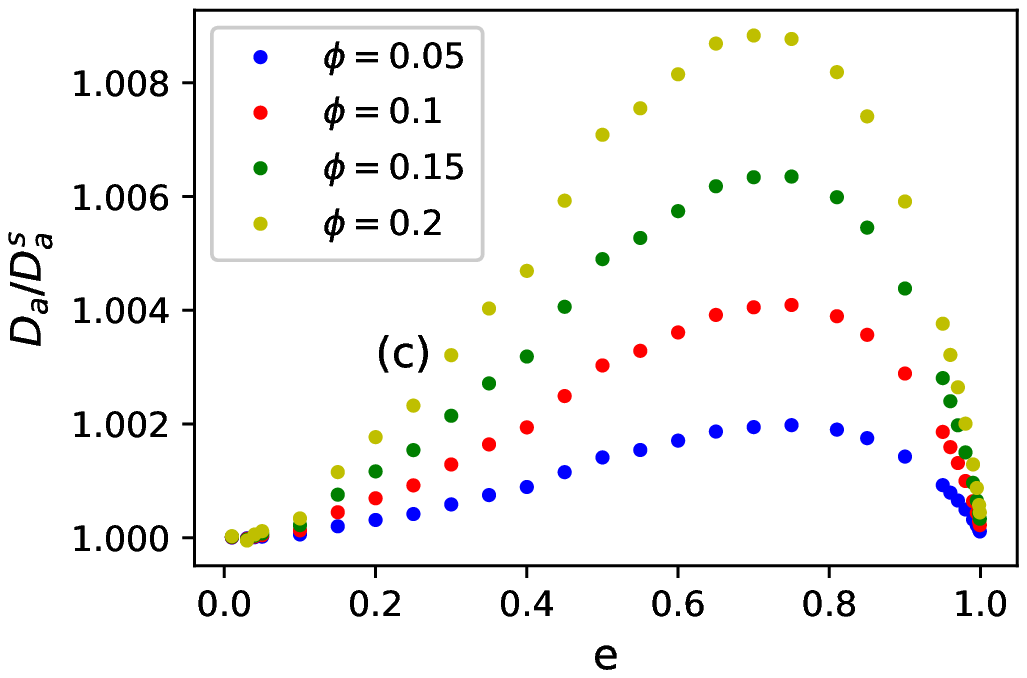}
\caption{\label{figurepercentp} (a) The hydrodynamic correction, $\phi \, D_{\perp}$ ($\phi \, D_{\parallel}$), in Eq.~(\ref{eq:finalanswer}) for a suspension of prolate spheroids ($\phi=0.1$) as a percentage of the single spheroid diffusivity $D_{\perp}^s$  ($D_{\parallel}^s$) versus the eccentricity. (b) $D_{a}$, the ratio defined in Eq.~(\ref{eq:ratio}), of a tracer spheroid in the suspension versus its eccentricity for various volume fractions. (c) The ratio scaled with its single particle value, $D_a^s$  ($=D_a|_{\phi=0}$), versus the eccentricity for various volume fractions. }
\end{figure}

\begin{figure}
\includegraphics[scale=.7]{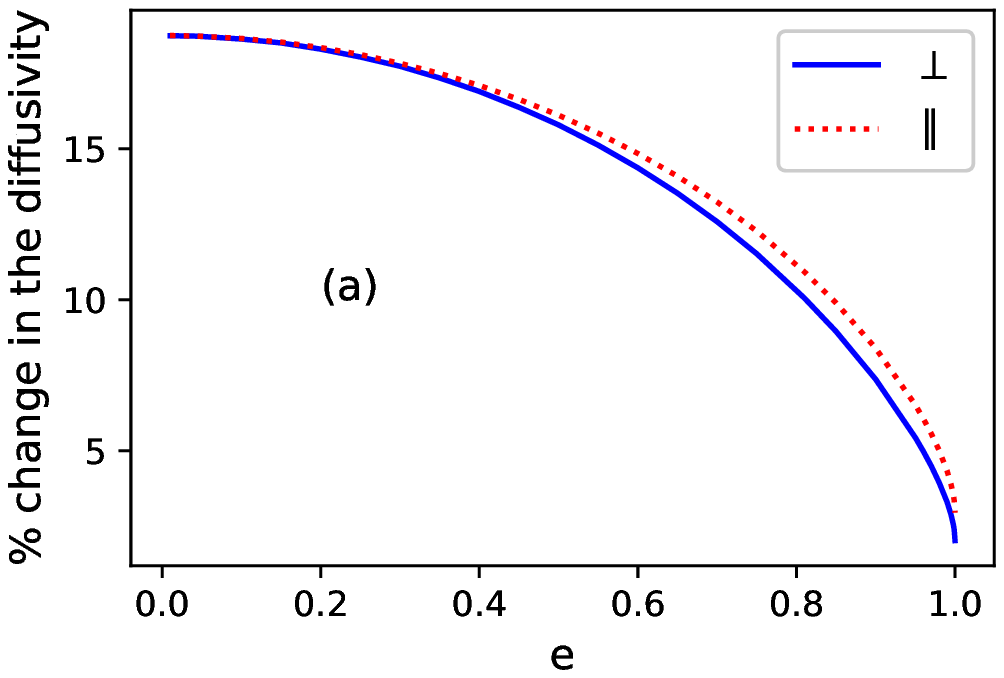}\\
\includegraphics[scale=.7]{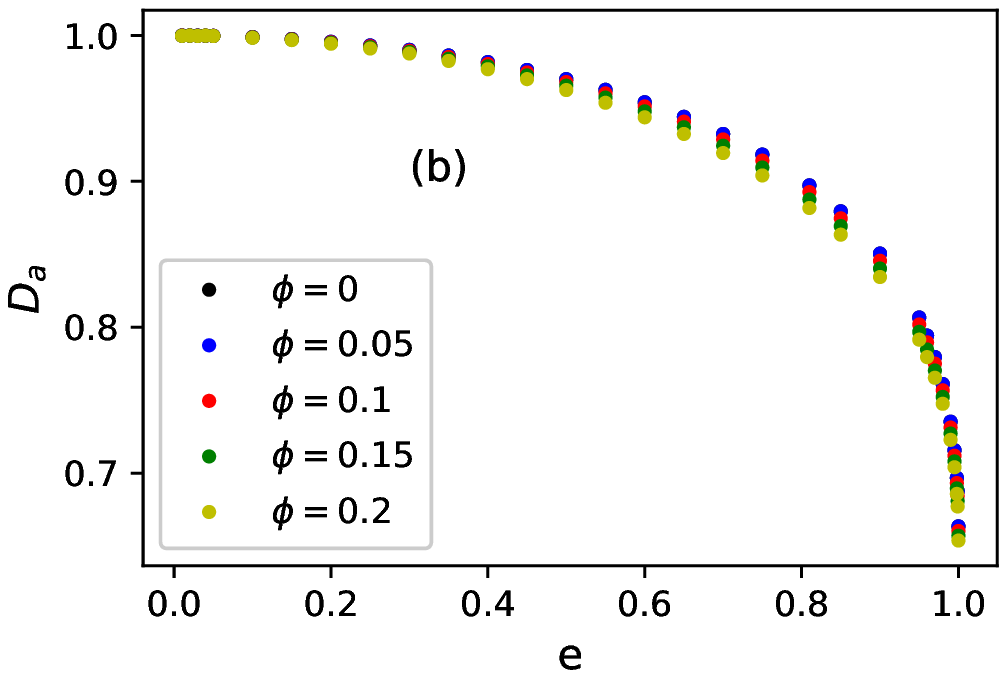}\\
\includegraphics[scale=.7]{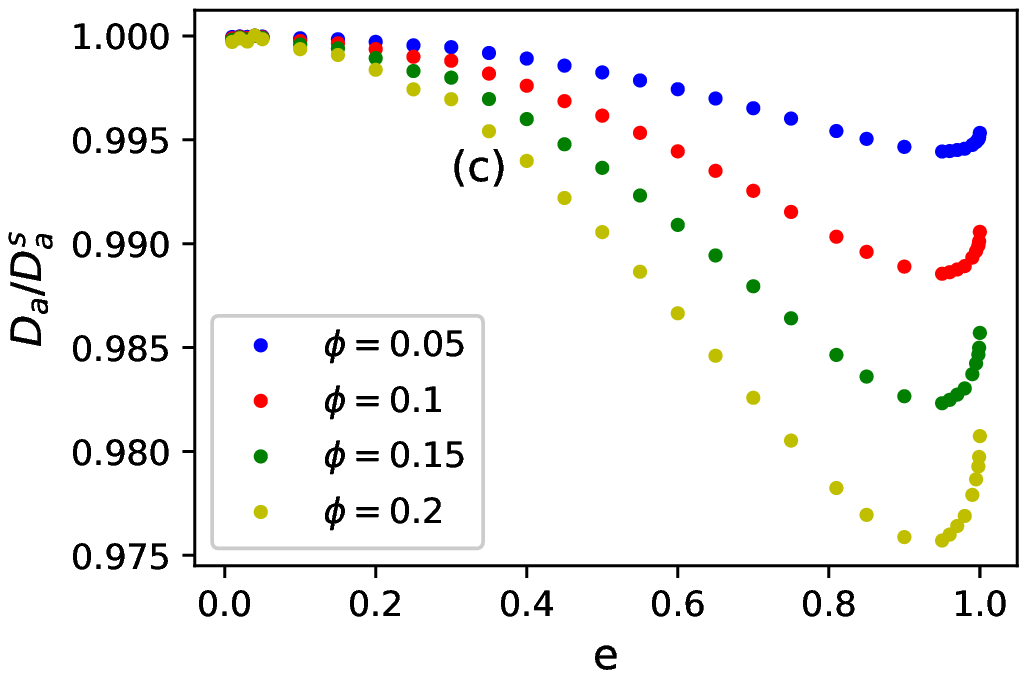}\\
\caption{\label{figurepercento} (a) The correction, $\phi \,D_{\perp}$ ($\phi \,D_{\parallel}$), in Eq.~(\ref{eq:finalanswer}) for a suspension of oblate spheroids ($\phi=0.1$) as a percentage of the single spheroid diffusivity $D_{\perp}^s$  ($D_{\parallel}^s$) versus the eccentricity. (b) $D_{a}$, the ratio defined in Eq.~(\ref{eq:ratio}), of a tracer spheroid in the suspension versus its eccentricity for various volume fractions. (c) The ratio scaled with its single particle value, $D_a^s$  ($=D_a|_{\phi=0}$), versus the eccentricity for various volume fractions. }
\end{figure}

\begin{figure}
\includegraphics[scale=.7]{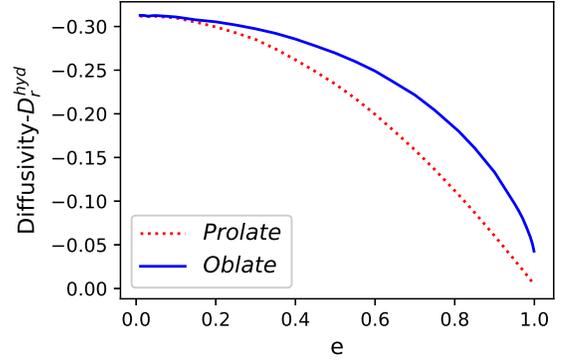}\\
\caption{\label{figurerot} The O($\phi$) hydrodynamic correction to the orientational diffusivity (transverse to a spheroid's axis) given in Eq.~(\ref{eq:finalanswer2}) versus eccentricity.}
\end{figure}
\begin{figure}
\includegraphics[scale=.7]{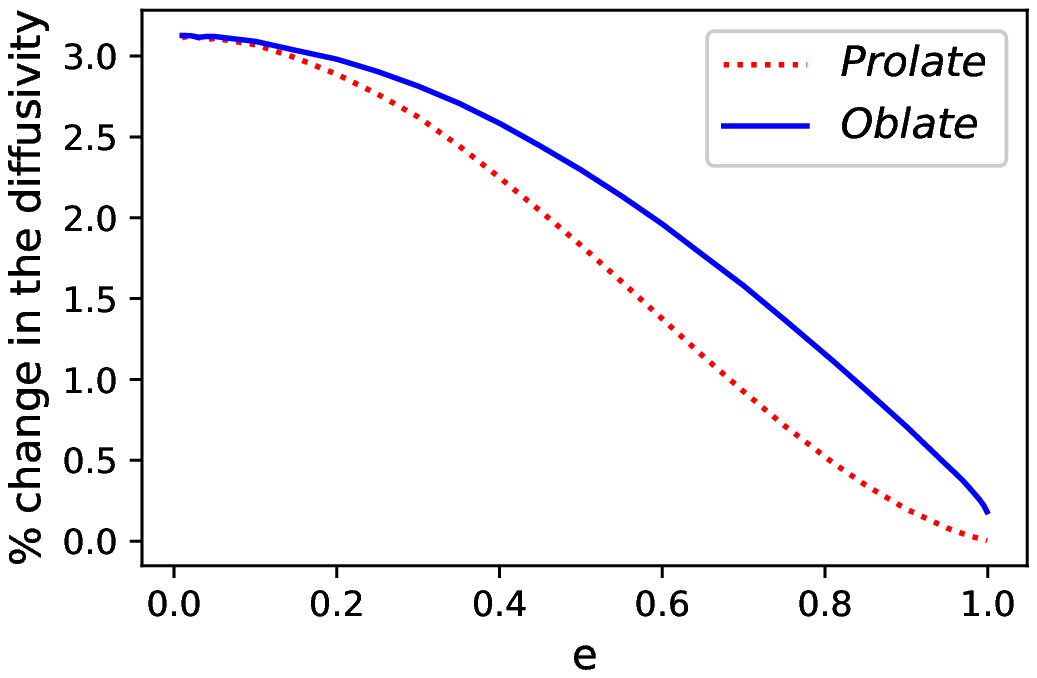}\\
\caption{\label{figurerot2} The correction to the orientational diffusivity, $\phi D_r^{\text hyd}$, (transverse to a spheroid's axis), expressed as a percentage of the associated single particle diffusivity, $D_r^s$, versus the eccentricity at $\phi=0.1$.}
\end{figure}

\section{\label{sec:Conclusions}Conclusions\\}
We have calculated the hydrodynamic corrections to the short-time translational and rotational diffusivities of a tracer spheroid of arbitrary aspect ratio in a dilute monodisperse suspension to O($\phi$). The corrections reduce the diffusivities of the spheroid from that of an isolated spheroid of the same aspect ratio. The corrections are monotonically decreasing functions of the spheroid eccentricity, peaking for a sphere, vanishing for a slender fiber, and asymptoting to finite values for a flat disk. 
\textcolor{black}{ Moreover, the corrections do not significantly alter the anisotropy of the translational diffusion.}

Further theoretical, numerical and experimental studies are needed to understand the behavior at higher particle volume fractions, where the lubrication effects neglected here become important. As $\phi$ increases, intuition suggests that the short-time diffusivities will be further reduced, since the reduction arises from the hydrodynamic hindrance created by the other particles in the suspension.

When one solidifies a colloidal suspension several processes relevant to the work described here are of interest.  Firstly, as noted in \S \ref{sec:level1}, the solid/suspension interface can become morphologically unstable owing to constitutional supercooling, the degree of which is controlled by the diffusivity of particles in the suspension.  Thus, our finding here of the suppression of the diffusivity implies that in natural settings, wherein the colloids are rarely spherical, morphological instability will be enhanced.  Moreover, because one observes a van't Hoff law in colloidal suspension\cite{peppin2008experimental}, our finding provides an underpinning for studies of its origin.  Secondly, once particles are trapped in ice they are surrounded by a thin unfrozen (``premelted'') water film that allows for their subsequent motion by thermomolecular pressure gradients\cite{JSW2019}.  The mobility of those particles can be treated within the framework of Onsager reciprocity\cite{Peppin:2009fe}, suggesting a range of interesting generalizations using spheroids, and in particular the question of thermodynamic buoyancy\cite{RWW:2001}.  Moreover, because of the temperature dependence of the viscosity of supercooled water \cite{dehaoui2015viscosity}, these dynamical processes will be influenced in a variety of ways, one of which concerns the temperature dependence of the ordering effect of water in confined supercooled films\cite{PW:2017}.  }

\begin{acknowledgments}
The authors acknowledge the support of the Swedish Research Council grant no. 638-2013-9243, conversation with their colleagues in Stockholm and Cristobal Arratia for access to his workstation on which the numerical integration code was executed.  
\end{acknowledgments}

\appendix

\section{\label{app:fdt} Derivation of the fluctuation dissipation theorem}
In this section, we derive the fluctuation-dissipation theorem (FDT) for a suspension of  $N$ spheroids.  One starts with Eq.~(\ref{eq:langeq}), albeit with an unknown relationship between the strength of the fluctuations and the mobility matrix, and noting that the time differentiation and ensemble average (over rapidly fluctuating random forces and torques for a given configuration of $N$ spheroids) commute, one can write,
\begin{equation}
\frac{d}{dt} \langle \mathbf{X} (t) \mathbf{X} (t)   \rangle = 2 \langle  \mathbf{X} (t) \mathbf{\dot{X}} (t) \rangle .  \label{eq:ensembderiv}
\end{equation}
As opposed to Eq.~(\ref{eq:forcecorr}), Eq. (\ref{eq:ensembderiv}) does not hold when the noise is $\delta$-autocorrelated because $\langle  \mathbf{X} (t) \mathbf{\dot{X}} (t) \rangle$ is ill-defined, although it does hold when the noise has a finite correlation time, a point to which we return in this derivation. 

For $t>t_m$, the angular and translational velocities of the spheroids equilibrate, and  hence $\frac{d}{dt} \langle \mathbf{X} (t) \mathbf{X} (t)\rangle=0$. Contracting Eq.~(\ref{eq:langeq}) with ${M}^{-1}_{li} X_f$ and averaging gives,
\begin{eqnarray}
    &\langle \dot{X}_l(t) {X}_f(t) \rangle   -   (I_s-I_t)   M_{li}^{-1}  \langle X^{I}_{i}(t) {X}_f(t)\rangle =- k_b T M_{li}^{-1} D_{ij}^{-1}\nonumber \\ 
 &\langle  X_j (t)  X_f (t)  \rangle +    M_{li}^{-1} {\alpha}_{ij}  \langle  g_j (t)  X_f (t) \rangle \label{eq:ensembderiv2}. 
\end{eqnarray}
Using Eq.~(\ref{eq:ensembderiv}), the first term of Eq.~(\ref{eq:ensembderiv2}) is zero, and the second term vanishes as it is an odd-moment in $\mathbf{X}$, and thus 
\begin{eqnarray}
 k_b T D_{ij}^{-1} \langle  X_j (t)  X_f (t)  \rangle- {\alpha}_{ij}  \langle  g_j (t)  X_f (t) \rangle =0 \label{eq:finfd}.
\end{eqnarray}
From the equipartition theorem \cite{dhont1996introduction}, the average in the first term above is equal to $k_b T M_{jf}^{-1}$. The second term is evaluated by noting that $\mathbf{X}$ is a functional of $\mathbf{g}$, namely $\mathbf{X}={{F}}[\mathbf{g}]$.  Therefore, $\langle  g_j (t)  X_f (t) \rangle$=$\langle  g_j (t) F_f[\mathbf{g}]  \rangle$ and because $\mathbf{g}$ is Gaussian noise, one finds\cite{zinn1996quantum}, 
\begin{eqnarray}
 \langle  g_j (t) F_f[\mathbf{g}]  \rangle= \int_0^t dt' \langle g_j (t) g_k (t')\rangle \left \langle  \frac{\delta F_f[\mathbf{g}]}{\delta g_k(t')} \right \rangle \label{eq:finfd2}.
\end{eqnarray}
The first average in the integrand above is given by Eq.~(\ref{eq:forcecorr}), and the second average is evaluated by solving Eq.~(\ref{eq:langeq}) to obtain the velocity vector (${\mathbf{X}}$), and then differentiating the vector with respect to $\mathbf{g}$. The average is given as
\begin{eqnarray}
 \left\langle \frac{\delta F_f[\mathbf{g}]}{\delta g_k(t')}\right\rangle&= \displaystyle\left\langle \frac{\delta }{\delta g_k(t')}\left(\int_{0}	^t [
 M^{-1}_{fl}  (I_s -I_t)X_l^{I} \right.\right.\nonumber \\ &\left.\left.- k_b T M^{-1}_{fl} D^{-1}_{lj} X_j ]dt'\vphantom{\displaystyle\int_{0}^{t}}\right) \right. \nonumber \\ &\left.+ \displaystyle  \int_{0}^t dt''  \delta(t'-t'') M^{-1}_{fl} {\alpha}_{lp}\delta_{pk}\right\rangle \nonumber\\
 &= \displaystyle\left\langle \frac{\delta }{\delta g_k(t')}\left(\int_{t'}^t [
 M^{-1}_{fl}  (I_s -I_t)X_l^{I}\nonumber \right.\right.\\ &\left.\left.- k_b T M^{-1}_{fl} D^{-1}_{lj} X_j ]dt'\vphantom{\displaystyle\int_{0}^{t}}\right) \right. \nonumber \\ &\left.  + \displaystyle  H(t-t') M^{-1}_{fl} {\alpha}_{lp}\delta_{pk}\vphantom{\left(\int_{0}^t\frac{1}{2}\right)}\right\rangle \label{eq:finfd4}, 
 \end{eqnarray}
where $H(t)$ is the Heaviside function. The $\delta$-autocorrelation of the noise in the integrand of Eq.~(\ref{eq:finfd2}) implies taking the limit of $t'\rightarrow t$ in  Eq.~(\ref{eq:finfd4}). In this limit, the contribution to Eq.~(\ref{eq:finfd2}) from the first term in Eq.~(\ref{eq:finfd4}) vanishes, and hence  Eq.~(\ref{eq:finfd}) becomes 
\begin{eqnarray}
 (k_b T)^2 D_{ij}^{-1} M^{-1}_{jf} =  2 {\alpha}_{ij} H(0) \alpha_{pj}M^{-1}_{pf}  \label{eq:finfd6}.
\end{eqnarray}
The Heaviside function at the origin, $H(0)$, is however discontinuous (often defined as the integral of a Dirac $\delta$-function) and arises from the fact that $\langle  \mathbf{X} (t) \mathbf{\dot{X}} (t) \rangle$ is ill-defined for $\delta$-autocorrelated noise. A solution to this problem is to instead use noise that has a finite autocorrelation time, viz.,  $\langle g_i(t) g_j(t') \rangle =2 \delta_{ij} \, \eta (t-t')$, where $\eta$ is an even function about $t=0$ such that $\int_{-\infty}^{\infty} \eta (t) dt=1$, 
and repeat the steps in this section up to Eq.~(\ref{eq:finfd6}). The net effect of using this colored noise is equivalent to replacing $H(0)$ with $\frac{1}{2}$ in Eq.~(\ref{eq:finfd6}), which leads to the FDT given in Eq.~(\ref{eq:flucdissi}). Note that, although we have derived the FDT for $N$ spheroids, this derivation can easily be extended to  $N$ rigid bodies.

\section{\label{app:subsec2} Derivation of the mean squared displacements}
In this section, starting from the Langevin equation (Eq.~\ref{eq:langeq}), we derive the translational and angular mean squared displacements of a tracer spheroid in a suspension of $N$ spheroids at short-times ($t_{m}\ll t \ll t_{c}$). The strength of the noise term in Eq.~(\ref{eq:langeq}) depends on the configuration of the particle described by  Eq.~(\ref{eq:flucdissi}). Hence, the governing equation for the positions and the orientations of the spheroids will be driven by a multiplicative noise. Although, such a noise leads to the well-known It\^{o}-Stratonovich dilemma\cite{moon2014interpretation}, the mean squared displacements at short-times ($\propto t$) required here, are unaffected by the calculus that is used. Eq.~(\ref{eq:langeq}) can be rewritten as
\begin{eqnarray}
{\dot{X}}_l+k_{b} T\, M^{-1}_{li} D_{ij}^{-1} X_j= (I_s-I_t) M^{-1}_{li} X_i^{I}  +   M^{-1}_{li}{\alpha}_{ij} g_j. \label{eq:langeq2}
\end{eqnarray}
Defining $Q_{lj}=k_{b} T\, M^{-1}_{li} D_{ij}^{-1}$, and noting that for times such that $t_{m}\ll t \ll t_{c}$, $\mathbf{Q}$ is independent of time, Eq.~(\ref{eq:langeq2}) can be written as
\begin{eqnarray}
\frac{d}{dt}\left(e^{Q t}_{ml} X_l\right)= e^{Q t}_{ml} \left[(I_s-I_t) M^{-1}_{li} X_i^{I}  +   M^{-1}_{li}{\alpha}_{ij} g_j\right]. \label{eq:langeq3}
\end{eqnarray}
One can integrate Eq.~(\ref{eq:langeq3}) twice with respect to time to find the translational and angular displacements of any spheroid. Since finite angular displacements are not vectors\cite{goldstein2011classical}, integrating the angular velocities with respect to time results in the correct angular displacements only when the displacements are infinitesimal, which is true here since $t \ll t_{c}$. The final translational displacement ($\Delta\mathbf{ r}$) of a tracer spheroid (say 1) in the suspension is given as 
\begin{eqnarray}
\Delta r_p^{1} =& \displaystyle \int_{0}^{t} Q^{-1}_{pq}\left[\delta_{ql}-e^{-Q (t-t')}_{ql}\right] M^{-1}_{li} \alpha_{ij} g_j dt'\nonumber \\
&+\displaystyle \int_{0}^{t} Q^{-1}_{pq}\left[\delta_{ql}-e^{-Q (t-t')}_{ql}\right](I_s-I_t)M^{-1}_{li}X_i^{I}dt' \label{eq:deltar}  \nonumber \\
&+ Q^{-1}_{pq}\left[\delta_{qm}-e^{-Q t}_{qm}\right]X_m(0),
\end{eqnarray}
where the integral containing $\mathbf{g}$ leads to a mean squared displacement that is proportional to $t$. The other two integrals lead to mean squared displacements that are smaller for short times, and hence are not considered. Thus, after neglecting those two integrals, the final mean squared displacement of the tracer spheroid is obtained by averaging the outer product $\Delta r_p^{1}  \Delta r_q^{1}$, first over an ensemble of realizations of random forces and torques, and then over an ensemble of various configurations of the spheroids in which the tracer particle is fixed in space. Using Eqs.~(\ref{eq:forcecorr}) and (\ref{eq:flucdissi}), one finds
\begin{eqnarray}
\langle\Delta r_p^1  \Delta r_q^1 \rangle= 2 \langle D^{tt}_{pq} \rangle t \label{eq:fintrans},
\end{eqnarray}
for the translational mean squared displacement of the tracer spheroid. The angular mean squared displacement of the spheroid is obtained similarly and is given by
\begin{eqnarray}
\langle\Delta \theta_p^1  \Delta \theta_q^1 \rangle= 2 \langle D^{rr}_{pq} \rangle t\label{eq:finang}.
\end{eqnarray}

\section{\label{app:mobility} Derivation of the mobility matrix for two prolate spheroids}
Using the method of reflections\cite{kim1985sedimentation,kim2013microhydrodynamics}, we derive the mobility matrix for two prolate spheroids that are subjected to forces as well as torques. 
 Consider two identical prolate spheroids that are subjected to torques and forces as shown in figure \ref{figure1}. The two spheroids will disturb the flow around them, and thereby interact with each other hydrodynamically. The aim is to calculate the relationship between the translational and the angular velocities of the spheroids and the forces and the torques. In the present calculation, the method of reflections has to be carried out up to the second reflection to obtain the leading order effects of hydrodynamic interactions. The translational ($\mathbf{U}_1$) and the angular velocities ($\boldsymbol{\omega}_1$) of the first spheroid (with orientation $\mathbf{p}_1$ in figure \ref{figure1}) are obtained by summing the contributions from different reflections that are given as
\begin{eqnarray}
 \mathbf{U}_1=& \mathbf{U}_1^{(0)}+\mathbf{U}_1^{(1)}+\mathbf{U}_1^{(2)},\label{eq:finmob0}
 \end{eqnarray}
 and
 \begin{eqnarray}
 \boldsymbol{\omega}_1=&\boldsymbol{\omega}_1^{(0)}+\boldsymbol{\omega}_1^{(1)}+\boldsymbol{\omega}_1^{(2)}, \label{eq:finmob00}
\end{eqnarray}
where the superscripts on the right-hand side indicate the number of reflections involved in obtaining that term.
 
At the zeroth reflection, the velocity field generated by a given spheroid in the figure is obtained using the method of singularities \cite{chwang1974hydromechanics,chwang1975hydromechanics}. The translational and the angular velocities of the first spheroid as well as the disturbance velocity field ($\mathbf{v}_1'$) at a point $\mathbf{x}$ generated by it \cite{kim1985sedimentation,claeys1993suspensions} are given as
\begin{eqnarray}
\mathbf{U}_1^{(0)}=& \displaystyle \frac{1}{6\pi \mu L}\left[\frac{1}{X_A} \mathbf{p}_1\,\mathbf{p}_1 + \frac{1}{Y_A} (\mathbf{I} -\mathbf{p}_1\, \mathbf{p}_1)\right]\mathbf{\cdot} \mathbf{F}_1,\label{eq:finmob1}
\end{eqnarray}
\begin{eqnarray}
\boldsymbol{\omega}_1^{(0)}=& \displaystyle \frac{1}{8\pi \mu L^3}\left[\frac{1}{X_C} \mathbf{p}_1\,\mathbf{p}_1 + \frac{1}{Y_C} (\mathbf{I} -\mathbf{p}_1\, \mathbf{p}_1)\right]\mathbf{\cdot} \mathbf{T}_1
\label{eq:finmob2},
\end{eqnarray}
and
\begin{eqnarray}
 \mathbf{v}_1'(\mathbf{x})=& \displaystyle \frac{\mathbf{F}_1}{16 \pi \mu c} \boldsymbol{\cdot} \int_{-c}^{c}\left[1+ (c^2-\xi_1^2)\frac{1-e^2}{4e^2}\nabla^2\right] \mathbf{J}(\mathbf{x}-\boldsymbol{\xi}_1) d\xi_1\label{eq:veldist}\nonumber\\
 &-\displaystyle \frac{3 }{64\pi\mu c^3}\,\mathbf{T}_1\,\boldsymbol{\cdot}\,\int_{-c}^{c}(c^2-\xi_1^2) \nabla \wedge \mathbf{J}(\mathbf{x}-\boldsymbol{\xi}_1)d\xi_1.\nonumber\\ \label{eq:mobility1}
\end{eqnarray}
Here, $c$  and $e$ are the semi interfocal distance and the eccentricity of the spheroid respectively, $\boldsymbol{\xi}_1= {\xi}_1 \mathbf{p}_1$ and $\mathbf{J}(\mathbf{x}-\boldsymbol{\xi}_1)= \frac{I}{|\mathbf{x}-\mathbf{\xi}_1|}+\frac{(\mathbf{x}-\mathbf{\xi}_1)(\mathbf{x}-\mathbf{\xi}_1)}{|\mathbf{x}-\mathbf{\xi}_1|^3}$ is the Oseen tensor\cite{kim2013microhydrodynamics}.  The quantities $X_A$, $Y_A$, $X_C$ and $Y_C$ are functions of the eccentricity \cite{kim2013microhydrodynamics} and are given at the end of this Appendix. The integrands in Eq.~(\ref{eq:mobility1}) are evaluated at all points between the foci from $\boldsymbol{\xi}_1= -c \mathbf{p}_1$ to  $c \mathbf{p}_1$. Eq.~(\ref{eq:mobility1}) implies that the velocity field due to the spheroid acted upon by a torque and a force is equivalent to that generated by a line distribution of singularities (Stokeslets, dipoles and degenerate quadrapoles) along the aforementioned points.

Since the force and the torque on the second spheroid (with orientation $\mathbf{p}_2$ in figure \ref{figure1}) are specified, $\mathbf{v}_1'$ cannot exert any additional force and torque upon it. Therefore, the spheroid  will rotate and translate such that the additional force as well as the torque are zero. However,  $\mathbf{v}_1'$ will induce a stresslet, $\mathbf{S}_2^{(1)}$, on the spheroid that will in turn lead to a reflected disturbance velocity field, $\mathbf{v}_{12}'$. The angular velocity, $\mathbf{\omega}_2^{(1)}$, and the stresslet  are obtained using Faxen's law \cite{kim1985note}, and are given by
\begin{eqnarray}
\mathbf{\omega}_2^{(1)}=&\displaystyle\frac{3}{4c^3} \frac{e^2}{2-e^2}\int_{-c}^{c}(c^2-\xi_2^2)\left[1+(c^2-\xi_2^2)\frac{1-e^2}{8 e^2 }\nabla^2\right]\nonumber\\& \displaystyle(\mathbf{p}_2\wedge \mathbf{e}_1'(\boldsymbol{\xi}_2) \boldsymbol{\cdot}\mathbf{p}_2)d \xi_2+\displaystyle \frac{3}{8c^3} \int_{-c}^{c} (c^2-\xi_2^2)\nabla \wedge \mathbf{v}_{1}' (\boldsymbol{\xi}_2) d \xi_2, \nonumber  \\
\end{eqnarray}
and
\begin{eqnarray}
 \mathbf{S}_2^{(1)}=&\displaystyle \mathbf{W}^{(2)}_{4} \boldsymbol{:} \frac{3}{4c^3}\int_{-c}^{c}(c^2-\xi_2^2)\left[1+(c^2-\xi_2^2)\frac{1-e^2}{8 e^2 }\nabla^2\right]\nonumber\\&\mathbf{e}_1'(\boldsymbol{\xi}_2) d\xi_2\ + \mathbf{Y}^{(2)}_{3} \boldsymbol{\cdot}\displaystyle\frac{3}{8c^3}\displaystyle\int_{-c}^{c}\nonumber  (c^2-\xi_2^2)\\&\left[\nabla \wedge \mathbf{v}_{1}' (\boldsymbol{\xi}_2)-2 \mathbf{\omega}_2^{(1)} \right]d\xi_2,
\end{eqnarray}\\
where $\boldsymbol{\xi}_2= {\xi}_2 \mathbf{p}_2$ changes from $-c\mathbf{p}_2$ to $c\mathbf{p}_2$ along the symmetry axis of the second spheroid. The tensor $\mathbf{e}_1'(\boldsymbol{\xi}_2)$ is the strain rate tensor due to $\mathbf{v}_1'(\boldsymbol{\xi}_2)$. The fourth order tensor $\mathbf{W}^{(2)}_{4}$ and the third order tensor $\mathbf{Y}^{(2)}_{3}$  are given by\cite{kim2013microhydrodynamics}

\begin{eqnarray}
 {W}^{(2)}_{4ijkl}=&\displaystyle\frac{20 \pi \mu L^3}{3} \left(X_M d_{ijkl}^{(0)}+Y_M d_{ijkl}^{(1)}+Z_M d_{ijkl}^{(2)}\right),
 \end{eqnarray}
 and
 \begin{eqnarray}
 {Y}^{(2)}_{3ijk}=&\displaystyle 4 \pi \mu L^3 Y_{H}\left(\epsilon_{ikl} p_{\scaleto{2j}{4pt}}+\epsilon_{jkl} p_{\scaleto{2i}{4pt}}\right)p_{\scaleto{2l}{4pt}},
\end{eqnarray}
where
\begin{eqnarray}
 d_{ijkl}^{(0)}=&\displaystyle\frac{3}{2}\left(p_{\scaleto{2i}{4pt}}p_{\scaleto{2j}{4pt}}-\delta_{ij}/{3}\right)\left(p_{\scaleto{2k}{4pt}}p_{\scaleto{2l}{4pt}}-\delta_{kl}/{3}\right),\\
  d_{ijkl}^{(1)}=&\displaystyle\frac{1}{2}\left(p_{\scaleto{2i}{4pt}}p_{\scaleto{2k}{4pt}}\delta_{jl}+p_{\scaleto{2j}{4pt}}p_{\scaleto{2k}{4pt}}\delta_{il}+p_{\scaleto{2i}{4pt}}p_{\scaleto{2k}{4pt}}\delta_{jl}+p_{\scaleto{2j}{4pt}}p_{\scaleto{2l}{4pt}}\delta_{ik}\nonumber\right. \\&\left.  - 4p_{\scaleto{2i}{4pt}}p_{\scaleto{2j}{4pt}}p_{\scaleto{2k}{4pt}}p_{\scaleto{2l}{4pt}}\right),
  \end{eqnarray} 
  and
\begin{eqnarray}  
   d_{ijkl}^{(2)}=&\displaystyle\frac{1}{2}\left(\delta_{ik} \delta_{jl}+ \delta_{jk} \delta_{il}-\delta_{ij} \delta_{kl}+\delta_{kl}p_{\scaleto{2i}{4pt}}p_{\scaleto{2j}{4pt}}+\delta_{ij}p_{\scaleto{2k}{4pt}}p_{\scaleto{2l}{4pt}} \right.\nonumber\\ & \left.  -p_{\scaleto{2i}{4pt}}p_{\scaleto{2k}{4pt}}\delta_{jl}-p_{\scaleto{2j}{4pt}}p_{\scaleto{2k}{4pt}}\delta_{il}-p_{\scaleto{2i}{4pt}}p_{\scaleto{2k}{4pt}}\delta_{jl}-p_{\scaleto{2j}{4pt}}p_{\scaleto{2l}{4pt}}\delta_{ik}\nonumber\right. \\&\left.  + p_{\scaleto{2i}{4pt}}p_{\scaleto{2j}{4pt}}p_{\scaleto{2k}{4pt}}p_{\scaleto{2l}{4pt}}\right).
\end{eqnarray}
The quantities $X_M$, $Y_M$, $Z_M$ and $Y_H$ are functions of the eccentricity and are given at the end of the section. The disturbance velocity field reflected by the second spheroid is given by
\begin{eqnarray}
 \mathbf{v}_{12}'(\mathbf{x})=& \displaystyle \mathbf{S}_2^{(1)}\boldsymbol{\cdot} \nabla \boldsymbol{\cdot} \frac{3}{4c^3} \int_{-c}^c(c^2-\xi_2'^2) \frac{1}{8\pi \mu} \\&\left[1+(c^2-\xi_2'^2)\frac{1-e^2}{8 e^2 }\nabla^2\right] \mathbf{J}(\mathbf{x}-\boldsymbol{\xi}_2') d\xi_2', \nonumber 
\end{eqnarray}
where $\boldsymbol{\xi}_2'= {\xi}_2' \mathbf{p}_2$ changes from $-c\mathbf{p}_2$ to $c\mathbf{p}_2$ along the symmetry axis of the second spheroid and the velocity field is proportional to $\mathbf{F}_1$ and $\mathbf{T}_1$, through $\mathbf{S}_2^{(1)}$. The velocity field cannot cause any additional torque and force on the first spheroid, and therefore the spheroid has to move with the translational and the angular velocities given by

\begin{eqnarray}
\mathbf{U}_1^{(2)}=& \displaystyle\frac{1}{2 c}\displaystyle \nonumber\int_{-c}^{c} \left[1+(c^2-\xi_1'^2)\frac{1-e^2}{4e^2} \nabla ^2\right] \mathbf{v}_{12}'(\boldsymbol{\xi}_1') d \xi_1' , \\ \label{eq:finmob3}
\end{eqnarray}
and
\begin{eqnarray}
 \mathbf{\omega}_1^{(2)}=&\displaystyle\frac{3}{4c^3} \frac{e^2}{2-e^2}\int_{-c}^{c}(c^2-\xi_1'^2)\left[1+(c^2-\xi_1'^2)\frac{1-e^2}{8 e^2 }\nabla^2\right]\nonumber\\& \displaystyle(\mathbf{p}_1\wedge \mathbf{e}_{12}'(\boldsymbol{\xi}_1') \mathbf{\cdot}\mathbf{p}_1)d \xi_1'+\displaystyle \frac{3}{8c^3} \int_{-c}^{c} (c^2-\xi_1'^2)\nabla \wedge \mathbf{v}_{12}' (\boldsymbol{\xi}_1') d \xi_1',  \nonumber \\ 
 \label{eq:finmob4}
 \end{eqnarray}
where $\boldsymbol{\xi}_1'= {\xi}_1' \mathbf{p}_1$ changes from $-c\mathbf{p}_1$ to $c\mathbf{p}_1$ along the symmetry axis of the first spheroid, and $\mathbf{e}_{12}'$ is the strain rate tensor due to  $\mathbf{v}_{12}'$. The Eqs.~(\ref{eq:finmob1}), (\ref{eq:finmob2}), (\ref{eq:finmob3}) and (\ref{eq:finmob4}) are substituted into Eqs.~ (\ref{eq:finmob0})and (\ref{eq:finmob00}) to obtain the elements of the mobility matrix that multiplies $\mathbf{F}_1$ and $\mathbf{T}_1$ to give $\mathbf{U}_1$ and $\boldsymbol{\omega}_1$ respectively. The velocities from the first reflection, $\mathbf{U}_1^{(1)}$ and $\boldsymbol{\omega}_1^{(1)}$ are not needed in the present calculation, as they are proportional to $\mathbf{F}_2$ and $\mathbf{T}_2$. Therefore, for a given configuration of the two spheroids in figure \ref{figure1}, the elements of the mobility matrix are  4-dimensional integrals over  $\xi_1$, $\xi_2$,  $\xi_1'$ and $\xi_2'$.\\

\begin{figure}
\includegraphics[scale=.5]{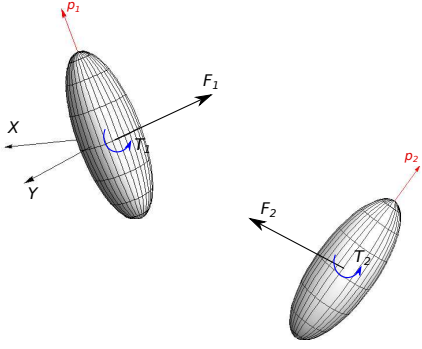}
\caption{\label{figure1} Two identical prolate spheroids with orientations $\mathbf{p}_1$ and $\mathbf{p}_2$  subjected to torques($\mathbf{T}_1$ and $\mathbf{T}_2$) and forces($\mathbf{F}_1$ and $\mathbf{F}_2$). XYZ is the coordinate system whose origin is at the centre of the first spheroid (with orientation $\mathbf{p}_1$) and $\mathbf{p}_1\parallel Z$.}
\end{figure}
The quantities needed to evaluate the mobility matrix are given by \cite{kim2013microhydrodynamics}
\begin{eqnarray}
X_A=\displaystyle\frac{8 e^3}{3\left[-2e+(1+e^2)\log\left(\frac{1+e}{1-e}\right)\right]}, \label{eq:appconst1}
\end{eqnarray}
\begin{eqnarray}
Y_A=\displaystyle\frac{16 e^3}{3\left[2e+(3e^2-1)\log\left(\frac{1+e}{1-e}\right)\right]},
\end{eqnarray}
\begin{eqnarray}
X_C=\displaystyle\frac{4 e^3(1-e^2)}{3\left(2e-(1-e^2)\log\left(\frac{1+e}{1-e}\right)\right)}\label{eq:appconst2},
\end{eqnarray}
\begin{eqnarray}
Y_C=\displaystyle\frac{4 e^3(2-e^2)}{3\left[-2e+(e^2+1)\log\left(\frac{1+e}{1-e}\right)\right]},
\end{eqnarray}
\begin{eqnarray}
X_M=\displaystyle\frac{8 e^5}{15\left[-6e+(3-e^2)\log\left(\frac{1+e}{1-e}\right)\right]} ,
\end{eqnarray}
\begin{eqnarray}
Y_M=&\displaystyle\frac{4 e^5}{5\left[-2e+(1+e^2)\log\left(\frac{1+e}{1-e}\right)\right]}\nonumber\\ &+ \displaystyle\frac{\left[2e(1-2e^2)-(1-e^2)\log\left(\frac{1+e}{1-e}\right)\right]}{\left[2e(2e^2-3)+3(1-e^2)\log\left(\frac{1+e}{1-e}\right)\right]},
\end{eqnarray}
\begin{eqnarray}
Z_M=\displaystyle\frac{16 e^5(1-e^2)}{5\left[-2e(3-5e^2)+3(1-e^2)^2\log\left(\frac{1+e}{1-e}\right)\right]},
\end{eqnarray}
and
\begin{eqnarray}
Y_H=\displaystyle\frac{4 e^5}{3\left[-2e+(1+e^2)\log\left(\frac{1+e}{1-e}\right)\right]}.
\end{eqnarray}\\

\section{\label{app:mobility2}Mobility matrix for two oblate spheroids}
Here we derive the mobility matrix for two oblate spheroids. The velocity field due to an oblate spheroid acted upon by a force and a torque is equivalent to that generated by a surface distribution of singularities (Stokeslets, dipoles and degenerate quadrapoles), the surface being a circle whose diameter is equal to the inter-focal length of the spheroid, and is on the equatorial plane of the spheroid with its centre being  the spheroid's centre of mass\cite{kim2013microhydrodynamics}. While one could derive the mobility matrix using the aforementioned surface distribution, in order to simplify the diffusivity calculation, the matrix is obtained by transforming that of the prolate spheroid derived in appendix \ref{app:mobility}. The transformation involves replacing the semi-inter-focal distance $c$ in the final expression for the velocities given in  Eqs.~(\ref{eq:finmob0}) and (\ref{eq:finmob00}) (after substituting for the velocities from Eqs.~(\ref{eq:finmob1}), (\ref{eq:finmob2}), (\ref{eq:finmob3}) and (\ref{eq:finmob4}))  with $-i\, c$, and replacing the eccentricity $e$ with $-i e/\sqrt{1-e^2}$, where the latter substitution means that the quantities $X_A,Y_A,Y_c,X_m, Y_m,Y_c, Y_H$ will be replaced by their oblate counterparts given at the end of this Appendix. The arguments leading to the transformation is given below.\\

The mobility matrix for the prolate spheroid was derived in Appendix \ref{app:mobility} using the singularity representation and Faxen's law for prolate spheroids. It is already known that the transformation, when applied to the singularity representation of the velocity field generated by a prolate spheroid of eccentricity $e$ in a particular flow (rotation, translation, straining, etc.,), gives the velocity field generated by an oblate spheroid of the same eccentricity in the same flow\cite{shatz2004singularity,dabade2015effects,dabade2016effect}. Now, the singularity representation can be interpreted as follows; the velocity field generated by a  spheroid is obtained by contracting a translational velocity (or angular velocity or strain rate) with a tensor obtained by operating a functional on $\frac{\mathbf{J}}{8\pi\mu}$, where $\mathbf{J}$ is the Oseen tensor.  For example, from Eq.~(\ref{eq:veldist}) the functional for translation would be $[6\pi\mu X_A \mathbf{p}_1\mathbf{p}_1+6\pi\mu Y_A(\mathbf{I}-\mathbf{p}_1\mathbf{p}_1)]({2 c})^{-1}\displaystyle\int_{-c}^{c}d\xi\left[1+(c^2-\xi^2)(1-e^2)/4e^2\nabla^2\right]$ , where the term in the first square bracket is the resistance tensor for a translating prolate spheroid. Applying the transformation on the singularity representation of the prolate spheroid is equivalent to applying it on this functional. The Lorentz reciprocal theorem can be used to show that the same functional, when used to operate on an ambient flow around a prolate spheroid, gives the driving force (or torque or stresslet)\cite{kim2013microhydrodynamics} due to the flow on the spheroid, leading to the Faxen law for prolate spheroids\cite{kim1985note}. Therefore, Faxen's law for oblate spheroids can be obtained from that of prolate spheroids by applying the transformation. Hence, by applying this transformation to the mobility matrix of a prolate spheroid of a given eccentricity, which was derived based on Faxen's law and the singularity representation, one can obtain that of an oblate spheroid of the same eccentricity.

The quantities needed to evaluate the mobility matrix for oblate spheroids are given by \cite{kim2013microhydrodynamics}.
\begin{eqnarray}
X_A=\displaystyle\frac{4 e^3}{3\left[(2e^2-1)\cot^{-1}\left(\frac{\sqrt{1-e^2}}{e}\right)+e\sqrt{1-e^2}\right]} \label{eq:appconst1},
\end{eqnarray}
\begin{eqnarray}
Y_A=\displaystyle\frac{8 e^3}{3\left[(2e^2+1)\cot^{-1}\left(\frac{\sqrt{1-e^2}}{e}\right)-e\sqrt{1-e^2}\right]} ,
\end{eqnarray}
\begin{eqnarray}
X_C=\displaystyle\frac{2 e^3}{3\left[\cot^{-1}\left(\frac{\sqrt{1-e^2}}{e}\right)-e\sqrt{1-e^2}\right]} \label{eq:appconst1},
\end{eqnarray}
\begin{eqnarray}
Y_C=\displaystyle\frac{2 e^3(2-e^2)}{3\left[(2e^2-1)\cot^{-1}\left(\frac{\sqrt{1-e^2}}{e}\right)+e\sqrt{1-e^2}\right]}\label{eq:appconst2},
\end{eqnarray}
\begin{eqnarray}
X_M=&\displaystyle\frac{4 e^5}{15\left(-3 e \sqrt{1-e^2}+(3-2 e^2)\cot^{-1}\left(\frac{\sqrt{1-e^2}}{e}\right)\right)} \label{eq:appconst3} ,
\end{eqnarray}
\begin{eqnarray}
Y_M=&\displaystyle\frac{2 e^5\left[e(1+e^2)-\sqrt{1-e^2}\, \cot^{-1}\left(\frac{\sqrt{1-e^2}}{e}\right)\right]}{5\left[3 e-e^3-3\sqrt{1-e^2}\cot^{-1}\left(\frac{\sqrt{1-e^2}}{e}\right)\right]}\nonumber\\ 
&\displaystyle\frac{1}{\left[e\sqrt{1-e^2}-(1-2e^2)\cot^{-1}\left(\frac{\sqrt{1-e^2}}{e}\right)\right]},
\end{eqnarray}
\begin{eqnarray}
Z_M=&\displaystyle\frac{8 e^5}{5\left[-(2 e^3+3e) \sqrt{1-e^2}+3\cot^{-1}\left(\frac{\sqrt{1-e^2}}{e}\right)\right]}\label{eq:appconst4},
\end{eqnarray}
and
\begin{eqnarray}
Y_H=&\displaystyle-\frac{2 e^5}{3\left[e \sqrt{1-e^2}-(1-2 e^2)\cot^{-1}\left(\frac{\sqrt{1-e^2}}{e}\right)\right]}.
\end{eqnarray}


\providecommand{\noopsort}[1]{}\providecommand{\singleletter}[1]{#1}%

\end{document}